\documentstyle[12pt, amsfonts, amssymb]{article}
\begin{document}



\def\be{\begin{equation}}
\def\ee{\end{equation}}
\def\ba{\begin{eqnarray}}
\def\ea{\end{eqnarray}}
\def\lb{\label}

\def\a{\alpha}
\def\b{\beta}
\def\g{\gamma}
\def\d{\delta}
\def\i{\eta}
\def\e{\varepsilon}
\def\l{\lambda}
\def\s{\sigma}
\def\t{\tau}
\def\r{\rho}
\def\v{\varphi}

\def\D{\Delta}
\def\G{\Gamma}
\def\L{\Lambda}
\def\P{\Phi}

\def\E{{\cal E}}
\def\cC{\cal C}

\def\fp{{\frak p}}

\def\C{\Bbb C}
\def\Z{\Bbb Z}
\def\F{\Bbb F}

\def\uz{\underline z}

\def\p{\hat p}

\def\bbr{{\rm I}\!{\rm R}}
\def\bbz{{\rm Z}\!\!\!{\rm Z}}

\def\Vp{{\cal V}_p}
\def\Hp{{\cal H}_p}

\def\bq{\overline{q}}

\def\id{\mbox{\rm 1\hspace{-3.5pt}I}}

\def\1{1\!\!{\rm I}}

\def\eod{\phantom{a}\hfill \rule{2.5mm}{2.5mm}}

\def\hF{\hat{F}}
\def\hA{\hat{A}}
\def\hB{\hat{B}}

\def\R{\hat{R}}
\def\Rp{\hat{R}(p)}

\def\uz{\underline z}
\def\pz{\Pi_{23}\;\uz}

\hyphenation{quad-ra-tic}

\newcommand{\I}[4]{
\left< {(p')^*}\left| \begin{array}{crcr} {#1} &{#2}\\
{#3} &{#4} \end{array} \right| p \right>}


\begin{center}


{\Large\bf Regular basis and R-matrices for the
}\\[3 mm]
{\Large\bf${\widehat{su}}(n)_k$ Knizhnik-Zamolodchikov
equation}\\[10 mm]


{
{\bf L.K. Hadjiivanov}}\footnote{e-mail address:
lhadji@inrne.bas.bg}\\ Theoretical Physics Division, Institute for
Nuclear Research and\\ Nuclear Energy, Tsarigradsko Chaussee 72,
BG-1784 Sofia, Bulgaria,\\
\vspace{2mm}
{
{\bf Ya.S. Stanev}}\footnote{On leave of absence from the
Institute for Nuclear Research and Nuclear Energy, BG-1784 Sofia,
Bulgaria; e-mail address: stanev@roma2.infn.it}\\ Dipartimento di
Fisica, Universit\`a di Roma "Tor Vergata",\\ I.N.F.N. -- Sezione
di Roma "Tor Vergata",\\ Via della Ricerca Scientifica 1, I-00133
Roma, Italy\\
\vspace{2mm}
and\\
\vspace{2mm}
{
{\bf I.T. Todorov}}\footnote{e-mail address:
todorov@inrne.bas.bg}\\ Theoretical Physics Division, Institute for
Nuclear Research and\\ Nuclear Energy, Tsarigradsko Chaussee 72,
BG-1784 Sofia, Bulgaria


\vspace{5mm}


\begin{abstract}

{\normalsize
\noindent
Dynamical $R$-matrix relations are derived for the group-valued
chiral vertex operators in the $SU(n)\,$ WZNW model from the KZ
equation for a general four-point function including two step
operators. They fit the exchange relations of the $U_q(sl_n)\,$
covariant quantum matrix algebra derived previously by solving the
dynamical Yang-Baxter equation. As a byproduct, we extend the
regular basis introduced earlier for $SU(2)\,$ chiral fields to
$SU(n)\,$ step operators and display the corresponding triangular
matrix representation of the braid group. }

\end{abstract}

\end{center}


\vspace{3mm}

\newpage


\textwidth = 16truecm
\textheight = 22truecm
\hoffset = -1truecm
\voffset = -2truecm


\section{Introduction}
\setcounter{equation}{0}
\renewcommand{\theequation}{\thesection.\arabic{equation}}

\medskip

There are, essentially, two different approaches to the
Wess-Zumino-Novikov-Witten (WZNW) model - a model describing the
conformally invariant dynamics of a closed string moving on a
compact Lie group $G\,$ \cite{W,GW}. The axiomatic approach
\cite{KZ} relies on the representation theory of Kac-Moody current
algebras and on the Sugawara formula for the stress-energy tensor.
The resulting chiral conformal block solutions of the
Knizhnik-Zamolodchikov (KZ) equation are multivalued analytic
functions which span a monodromy representation of the braid group
\cite{TK, K}. Surprisingly (at least at first sight), the
associated symmetry was related to the recently discovered quantum
groups \cite{A-GGS, FFK, FGP, T, Gaw, STH}. This relation was
explained in some sense by the second, canonical approach to the
problem \cite{B, F1, BDF, F2, G, FG, Chu, FHT1, FHT2, FHT3, CL,
FHIOPT}. The Poisson-Lie symmetry \cite{S-T-S} of the WZNW action
\cite{F1, F2, G} indeed gives rise to quantum group invariant
quadratic exchange relations \cite{F1, FG, AF} at the quantum
level.

In spite of continuing efforts \cite{DT, goslar, FHIOPT}, the
correspondence between the two approaches is as yet only
tentative: there is still no consistent operator formulation of
the chiral WZNW model to date that would reproduce the known
conformal blocks. The objective of the present paper is to provide a
step in filling this gap.

The first problem we are addressing is to find the precise
correspondence between the monodromy representation of the braid
group \cite{TK, K} and the $R$-matrix exchange relations
\cite{F1, F2, FG} among step operators -- i.e., field operators
transforming under the defining $n$-dimensional ("quark")
representation of $SU(n)\,.$ To this end we consider the four-point
"conformal block" of two step operators sandwitched between a pair
of primary chiral fields transforming under arbitrary {\em irreducible
representations} (IR) of $SU(n)\,,$ only restricted by the
condition that the 
resulting space of $SU(n)\,$ invariants 
is non-empty. In
fact, for a given initial and final states $|p{\cal i}\,$ and
${\cal h}p' |\,$ (labeled by the highest weights of $SU(n)\,$ IR
-- see Section 2), the space ${\cal F}(p, p' )\,$ of invariant
tensors of the type
\be
\lb{invt}
{\cal F}(p, p' ) = \{\,{\cal h}p' | \v_1^A\;\v_2^B   |p{\cal i}\,,\
A,B=1,\dots ,n\,\}
\ee
(the subscripts $1$ and $2$ replacing the world
sheet variables and discrete quantum numbers other than the $SU(n)\,$ indices
$A , B\,$) can be either $0, 1$ or $2$-dimensional. We
concentrate on the most interesting $2$-dimensional case that
includes the antisymmetric tensor product of $\v_1\,$ and
$\v_2\,$) but also write down the $1$-dimensional ("anyonic")
braid relations.

A new result of the paper is the extension to
${\widehat{su}}(n)\,$ step
operators of the "regular basis" (introduced originally for
${\widehat{su}}(2)\,$ blocks \cite{STH} as a counterpart of a
distinguished basis of quantum group, $U_q(sl_2)\,,\,$ invariants
\cite{FST}). Moreover, we demonstrate that the M\"obius invariant
amplitude for the ${\widehat{su}}(2)\,$ and the
${\widehat{su}}(n)\,$ theory coincide and so do the
"normalized braid matrices" $q^{\frac{1}{n}} B\,.$ This allows
to extend earlier results on the Schwarz finite monodromy problem
for the ${\widehat{su}}(2)\,$ KZ equation to the
${\widehat{su}}(n)\,$ case.

We start, in Section 2, with some background material including
various forms of the KZ equation. Special attention is devoted to
two bases of $SU(n)\,$ invariants (whose properties and
interrelations are spelled out in Appendix A). They appear as
prototypes of the $s$-channel basis and regular basis of solutions
of the KZ equation studied in Sections 3 and 4, respectively. The
standard notion of a chiral vertex operator (CVO) \cite{TK} and
its zero modes' counterpart \cite{AF, FG, Chu, FHT2, FHT3, CL,
FHIOPT} are applied in Section 3 for studying the braid properties
of the "physical solutions" of the KZ equation. It is the regular
basis introduced in Section 4 that is appropriate to also include
its logarithmic solutions.



\vspace{5mm}

\section{KZ equation for a $4$-point conformal block}
\setcounter{equation}{0}
\renewcommand{\theequation}{\thesection.\arabic{equation}}

\medskip

We label the IR of $SU(n)\,$ by their shifted highest weights
(see \cite{FHIOPT}):
\be
\lb{p}
p_{i\;i+1} = p_i-p_{i+1} = {\l}_i + 1\quad (\;{\l}_i \in {\Z}_+\;)\quad
{\rm where}\quad \sum_{i=1}^n p_i = 0\,.
\ee
Let the highest weight $p'\,$ belong to the tensor product of $p =
(p_1 , \dots , p_n )\,$ with a pair of "quark" IR (with $\l_i =
\d_{i1}\,$). The basic object of our study will be the four-point
block
\be
\lb{w}
w^{AB}_{\a\b} (\uz ; p,p' )\equiv
w^{AB}_{\a\b} (z_1 , z_2 , z_3 , z_4 ; p , p' ) =
{\cal h}0|
\P_{{p'}^*} (z_1) \v^A_\a (z_2) \v^B_\b (z_3) \P_p (z_4)
|0{\cal i}
\ee
where $\v^A_\a (z)\,$ is the fundamental chiral ("quark", or
"group valued") field ($A\,,\ \a\,$ are $SU(n)\,$ and
$U_q(sl_n)\,$ indices, respectively), $\P_p (z)\,$
is a (primary) chiral field carrying weight $p\,$
(whose tensor indices are omitted), and
${p'}^* = (-{p'}_n,\dots ,-{p'}_1)$ is the weight conjugate to
$p'\,.$ Let $v^{(i)}\,$ be the shift of weight under
the application of an $SU(n)\,$ step operator $\v^i_\a (z)\,:$
\be
\lb{vi}
( v^{(i)} | p ) = p_i\,,\quad (v^{(i)} | v^{(j)} ) = \d_{ij} -
\frac{1}{n}\,,\quad \sum_{i=1}^n v^{(i)} = 0\,.
\ee
Then  $p'\,$ and $p\,$ satisfy
\be
\lb{p'-p}
p' - p = v^{(i)} + v^{(j)}\equiv
v^{(m)} + v^{(m')}\,,\quad
m={\rm min}\, (i,j)\,,\
m' ={\rm max}\, (i,j)
\ee
where we assume that $p\,, p'\,$ and $p+v^{(m)}\,$
are dominant weights.

We shall express the four-point block $w\,$ (\ref{w}) in terms of a
conformally invariant amplitude $F\,$ setting
\be
\lb{w=DF}
w^{AB}_{\a\b} (\uz ; p,p' ) \,=\, D(\uz ; p,p' )\,
F^{AB}_{\a\b} (\i ; p,p' )
\ee
where the cross ratio $\i\,$ and the prefactor $D(\uz ; p,p' )\,$
are given by
\ba
\lb{D}
&&\i = \frac{z_{12}z_{34}}{z_{13}z_{24}}\,,\quad z_{ij}=z_i-z_j\,,\\
&&D (\uz ; p,p' ) = \left({{z_{24}}\over{z_{12}z_{14}}}\right)^{\D (p')}
\left({{z_{13}}\over{z_{14}z_{34}}}\right)^{\D (p)}
z_{23}^{-2\D}\ \i^{{\D}' - \D}\, (1-\i )^{\D_{(a)}}\,.\nonumber
\ea
Here
\ba
\lb{DD}
&&2 h \D (p) \equiv C_2(p) =\frac{1}{n} {\sum_{r<s}p_{rs}^2}-
\frac{n(n^2-1)}{12}\,,\quad\D = \frac{n^2-1}{2nh}\,,\\
&&{\D}'\equiv\D (p+v^{(m')}) =
\D (p)+\frac{p_{m'}}{h}+\frac{n-1}{2nh}\,,\quad
\D_{(a)} = \frac{(n+1)(n-2)}{nh}\nonumber
\ea
(For the current algebra ${\widehat{su}}(n)_k\,$ the {\em height}
$h\,$ is given by $h=k+n\,,\ k\,$ being the level.) Note that
$$
2 p_{m'} = h \left( \D (p') - \D (p) \right) - (p_{m m'}+\d_{m m'})
+\frac{2-n}{n}\,.
$$
\vspace{5mm}

\noindent
{\bf Remark 2.1~} The prefactor, a product of powers of the
differences $z_{ij}\,,$ is determined
by the overall scale dimension of $w^{AB}_{\a\b} (\uz ; p,p' )\,$ and
by infinitesimal $L_1\,$
invariance, i.e.,
\ba
\lb{M}
&&\sum_{a=1}^4 \;z_a^\nu (z_a {{\partial}\over{\partial
z_a}} + (\nu +1)\;\D_a )\; D (\uz ; p,p' ) = 0\quad{\rm for}
\quad\nu = 0,\pm 1\,,\nonumber\\
&&\D_1 = \D ({p'}^*) \equiv \D (p' )\,,\quad
\D_2 = \D_3 = \D\,,\quad
\D_4 = \D (p)
\ea
(see (\ref{DD})), up to a monomial in $\i\,$ and $1-\i\,.$ Our choice
(\ref{D}) corresponds to extracting the leading singularities so that
$F^{AB}_{\a\b} (\i ; p, p' )\,$ should be finite and nonzero at both
$\i =0\,$ and $\i =1\,.$

\vspace{5mm}

Let $C_{ab} = {\bf t}_a .{\bf t}_b\,$ be the polarized Casimir invariant
($a,b = 1,\dots ,4$). The generators ${\bf t}_a\,$ of the
representation $(a)\,$ of $SU(n)\,$ are normalized in such a way
that if $(a)\,$ refers to an IR of weight $p\,,$ then ${\bf
t}_a^2\,$ coincides with $C_2(p)\,$ of (\ref{DD}); in our case,
with $C_a := {\bf t}_a^2\,,$ we have
\be
\lb{C2}
C_4\, (\; ={\bf t}_4^2\;)\, = C_2 (p) =
\frac{1}{n}\sum_{i<j}\left( p_{ij}^2-(j-i)^2\right)=\frac{1}{n}
\sum_{i<j} p_{ij}^2 -\frac{n(n^2-1)}{12}\,.
\ee
$SU(n)\,$ invariance of the Wightman function implies
\be
\lb{suninv}
\left({\bf t}_1+{\bf t}_2+{\bf t}_3+{\bf t}_4 \right)
w_{\a\b}(\uz ; p,p' ) = 0 =
\left( C_a + \sum_{b\ne a} C_{ab}\right)
w_{\a\b}(\uz ; p,p' )\,.
\ee
The KZ equation
\be
\lb{KZ}
\left( h{{\partial}\over{\partial z_a}} - \sum\limits_{b\ne a}\;
{{C_{ab}}\over{z_{ab}}} \right)\;
w_{\a\b} (\uz ; p,p' )\; =\; 0\,,\quad 1\le a\le 4\,,
\ee
yields (choosing, say, $a=2\,$ and using (\ref{suninv}))
\be
\lb{KZeta}
\left( h\frac{d}{d\i} - \frac{\Omega_{12}}{\i} + \frac{\Omega_{23}}{1-\i}
\right) F_{\a\b} (\i ; p,p' ) = 0
\ee
where
\be
\lb{Omega}
\Omega_{12} = C_{12} + p_m + \d_{m m'} + \frac{n^2+n-4}{2n}\,,\quad
\Omega_{23} = C_{23} + \frac{1}{n} + \id = P_{23} + \id\,,
\ee
$P_{23}\,$ being the permutation operator for the two factors in
the tensor product ${\C}^n\otimes{\C}^n\,$ of fundamental IR
($P_{23}^2 = \id\;$).
$\Omega_{12}\,$ and $\Omega_{23}\,$ satisfy
\ba
\lb{OOO}
&&\Omega_{12}\,\Omega_{23}\,\Omega_{12}\, =\, ({\frak p}-1)\,
\Omega_{12}\,,\quad
\Omega_{23}\,\Omega_{12}\,\Omega_{23}\, =\, ({\frak p}-1)\,
\Omega_{23}\\
&&{\frak p}\, =\, p_{m m'} + \d_{m m'}\,\in\,{\Bbb N} \,,\qquad
\Omega_{12}^2\, =\, p_{m m'}\, \Omega_{12}\,,
\qquad \Omega_{23}^2\, =\, 2\, \Omega_{23}\nonumber
\ea
-- see Appendix A.

The value $\fp =1\,$ is special and we shall consider it separately.

For $\fp \ge 2\,$ both
$\Omega_{12}\,$ and $\Omega_{23}\,$ are nontrivial (nonnegative)
operators with a single eigenvalue zero. In this case the space
${\cal F}(p,p')\ (\;\ni F^{AB}\; )\,$
is conveniently spanned by the eigenvectors $I_0 = (I_0^{AB} )\,$
and $I_1 = (I_1^{AB} )\,$ of $\Omega_{12}\,$ and $\Omega_{23}\,,$
respectively, corresponding to eigenvalue $0\,$ (cf. \cite{STH}
and Appendix A below):
\be
\lb{I}
\Omega_{12} I_0 = 0 = \Omega_{23} I_1\,,\quad I_1 = (P_{23}-1 ) I_0\,.
\ee
We shall set
\be
\lb{F=If}
F^{AB}_{\a\b} (\i ;p,p') =
I_0^{AB}\;(1-\i )\; f^0_{\a\b} (\i ) +
I_1^{AB}\; \i\; f^1_{\a\b} (\i )\ .
\ee
Inserting (\ref{F=If}) into (\ref{KZeta}), we find the following
first order system for
$f^{\ell}\,\equiv f^{\ell}_{\a\b} (\i )\,,\ \ell = 0,1\,$:
\be
\lb{system}
h (1-\i ) \frac{d f^0}{d \i} = (h-2) f^0 + ({\frak p}-1) f^1\,,\quad
h\;\i\;\frac{d f^1}{d \i} = ({\frak p} -h) f^1 - f^0\,.
\ee
It yields a hypergeometric (HG) equation for each $f^{\ell}\,,\
{\ell}=0,1\,:$
\be
\lb{HGE}
\i (1-\i ) \frac{d^2 f^{\ell}}{d{\i}^2} +
\left( 1+\ell -\frac{\frak p}{h} - (3-\frac{{\frak p}+2}{h} )\i
\right) \frac{d f^{\ell}}{d \i} = (1 -
\frac{1}{h})(1-\frac{{\frak p}+1}{h}) f^{\ell}\,.
\ee

For $\fp =1\,$ we have, in view of (\ref{OOO}), either $m=m'\,$
implying
\be
\lb{m=}
P_{23} S_0 = S_0 = I_0\,,\quad I_1 = 0\,,\quad S_1 = S_0\,,\quad
F (\i ) = K S_0 (1-\i )^{\frac{2}{h}} \,,
\ee
or $m' =m+1\,,\ \, p_{m\,m+1}=1\,,$ when
\be
\lb{m>}
P_{23} S_1 = - S_1\,, \quad S_0 = 0 = I_0 = I_1\,,\quad
F (\i ) = K S_1 \i^{\frac{1}{h}}
\ee
(see Appendix A). In both cases ${\rm dim}\, {\cal F} (p, p' ) \le
1\,.$ In particular, for $p = p^{(0)}\,,\ m = m'-1 = 1\,$ and
for $p' = p^{(0)}\,,\ m' = m+1 = n\,$ we are dealing with a
3-point function.

It is remarkable that the KZ equation (\ref{KZeta}) for the
M\"obius invariant amplitude $F_{\a\b}(\i ;p,p')\,$ is in fact
independent of the group label $n$ -- as made manifest by the
system (\ref{system}). Only the prefactor $D (\uz ;p,p')\,$
(\ref{D}) carries an $n$-dependence.

\vspace{5mm}

\section{$s$-channel basis of solutions. Braid relations}
\setcounter{equation}{0}
\renewcommand{\theequation}{\thesection.\arabic{equation}}

\medskip

The expansion of the basic fields $\v^A_\a (z)\,$ into {\em
chiral vertex operators} (CVO), $\v^A_\a (z) = \v^A_i (z)
a^i_\a\,$ \cite{FHT2, FHT3, FHIOPT} gives rise to an expansion of
$f^{\ell}_{\a\b}\,$ into $s$-channel conformal blocks. We have
\be
\lb{f=sS}
f^{\ell}_{\a\b}(\i ) = \sum_{{\l}=0}^1 s^{\ell}_{\l} (\i )
{\cal S}^{\l}_{\a\b}\,,\
{\cal S}^0_{\a\b} = {\cal h} p' | a^m_\a a^{m'}_\b |p{\cal i}\,,\
{\cal S}^1_{\a\b} = {\cal h} p' | a^{m'}_\a a^m_\b |p{\cal i}
\ee
for
\be
\v^A_\a (z)\; =\; c^A_i \; \v^i_j (z)\; a^j_\a\,.
\ee
Here $a^j_\a\,,$ (which, together with $q^{p_{ij}}\,,$
generate a {\em quantum} matrix algebra of $SL(n)\,$ type
\cite{HIOPT, FHIOPT}) satisfy
\setcounter{equation}{0}
\renewcommand{\theequation}{\thesection.3\alph{equation}}
\ba
&&a^i_\a a^j_\a = a^j_\a a^i_\a\,,\quad a^i_\a a^i_\b
=q^{\epsilon_{\a\b}} a^i_\b a^i_\a \quad{\rm and,\ for}\
i\ne j\,,\ \a\ne\b\,,\\
&&\rho (p_{ij})[p_{ij}-1]a^j_\a a^i_\b=
[p_{ij}] a^i_\b a^j_\a- q^{\epsilon_{\b\a} p_{ij}}
a^i_\a a^j_\b
\qquad\left(\rho (p)\rho (-p) = 1\right)\,.\nonumber
\ea
We are using the notation of Section 2.3 in \cite{FHIOPT}
$$\epsilon_{\b\a}=1=-\epsilon_{\a\b}\ {\rm for}\ \a <\b\,,\
\epsilon_{\a\a}=0\ ;\quad [p]:=\frac{q^p -\bq^p}{q-\bq}\,;
\quad q=e^{-i\frac{\pi}{h}}={\bq\,}^{-1}\,.
$$
The factor $\rho(p_{ij})$ constraint by the last equation in
(3.2a) reflects the "gauge freedom" in the solution of the
dynamical Yang-Baxter equation \cite{GN, I2, HIOPT, FHIOPT}.
The $SU(n)\,$ tensor operators $c^A_i\,$ are generators of
the undeformed counterpart of the matrix algebra (see Appendix
A); in particular,
\ba
&&[c^A_i , c^B_i ] = 0 = [c^A_i , c^A_j ] \,,\quad{\rm and,\ for}\
i\ne j\,,\ A\ne B\,,\\
&&r(p_{ij} ) \, ( p_{ij} - 1)\, c^A_j c^B_i =
 p_{ij} c^B_i c^A_j - c^A_i c^B_j \qquad
\left( r(p) r(-p) = 1\right)\,.\nonumber
\ea

\setcounter{equation}{3}
\renewcommand{\theequation}{\thesection.\arabic{equation}}
The function $s^{\ell}_0 (\i )\,$ is characterized as the analytic
around the origin solution of the system (\ref{system}), resp. of the
HG equation (\ref{HGE}) satisfying $K_0 := s^0_0 (0) > 0\,.$ It is
given by the HG series
\be
\lb{s00}
s^0_0 (\i ) = K_0\; F(\a ,\b ; 1-\a +\b ;\i )
= \frac{K_0}{B(1-\a , \b)}
\int_0^1 t^{\b -1}(1-t)^{-\a}(1-t\i )^{-\a} dt
\ee
where $B(x,y) = \frac{\G (x)\G (y)}{\G (x+y)}\,$ is the
beta-function, 
$p' -p\,$ is given by (\ref{p'-p}),
$p_{m m'} = {\frak p}\ge 2\,$
and
$\a = 1-\frac{1}{h}\,,\ \b = 1-\frac{{\frak p}+1}{h}\ \ 
(1-\a+\b = 1-\frac{\fp}{h} ) .\,$
For ${\frak p}=h-1\,,\ s^0_0 (\i ) = K_0\,$
($\lim\limits_{\b \to 0} \frac{t^{\b -1}}{\G (\b )}
= \d (t)\,$). From (\ref{system}) one gets
\be
\lb{s10}
s^1_0 (\i )
= \frac{\a -1}{1-\a +\b}\; K_0\; F (\a ,\b ; 2-\a +\b ;\i )\,.
\ee
The second solution, $s^{\ell}_1 (\i )\,,$ is obtained requiring
that
\be
\lb{sec}
\i^{\D (p+v^{(m' )}) -\D (p+v^{(m)}) } s^0_1 (\i ) =
\i^{-\frac{\frak p}{h}} s^0_1 (\i )
\ee
is analytic around $\i = 0\,;$ we find
\be
\lb{s1}
s^0_1 (\i ) = K_1\; \i^{\a -\b}
F (\a ,2\a -\b ; 1+\a -\b ;\i )\,,
\ee
\be
\lb{s11}
s^1_1 (\i ) = \frac{\a -\b}{2\a -1-\b} \; K_1\; \i^{\a -\b -1}
F (\a -1,2\a -1-\b ; \a -\b ;\i )\,.
\ee

We shall now derive the braid relation among the above conformal
blocks under the exchange of two step operators $\v\,.$

The four-point blocks
$w^{AB}_{\a\b} (z_1 , z_2 , z_3 , z_4 ; p,p' )\,$
(\ref{w=DF}) are single valued in the domain
\be
\lb{dom}
{\cal O}_4=\{ (z_1 , z_2 , z_3 , z_4)\in{\C}^4\,; |z_i |>
|z_{i+1}|\,,\, i=1,2,3\,; |{\rm arg} z_i |<\pi\,,\, i=1,2,3,4 \}
\ee
We consider the analytic continuation
$\stackrel{\curvearrowright}{{w}^{BA}_{\a\b}}
(z_1 , z_3 , z_2 , z_4 ; p,p' )$ of
\newline
$w^{BA}_{\a\b} (z_1 , z_2 , z_3 , z_4 ; p,p' )$
along paths in the homotopy class of a particular
curve
\setcounter{equation}{0}
\renewcommand{\theequation}{\thesection.10\alph{equation}}
\be
{\cal C}(2,3)\,:\;
[0, 1]\times \left({\C}^4\setminus \{ z_i=z_j\,, i\ne j\}\right)\;
\to\;{\C}^4\setminus \{ z_i=z_j\,, i\ne j\}
\ee
which interchanges $z_2\,$ and $z_3\,$ and whose internal points
belong to ${\cal O}_4\,$ whenever the end points belong to
its boundary(\ref{dom}); e.g., for
$z_i(t) = e^{i\zeta_i (t)}\,,\ i=1,2,3,4\,,$
\ba
&&\zeta_1 (0)=x_1 -i\epsilon\,,\
\zeta_{2,3} (0)=x_{2,3}\,,\
\zeta_4 (0)=x_4 +i\epsilon\,,\nonumber\\
&&x_i \in {\Bbb R}\,,\quad \epsilon >0\,,\quad
x_1 > x_2 > x_3 > x_4\,,\ \  x_{14}<\pi\,,
\ea
we define ${\cal C}(2,3)\,$ by
\be
\lb{homot}
\zeta_1(t)=\zeta_1\,,\ \zeta_{2,3}(t) =
e^{-i\frac{\pi}{2}t}(x_{2,3}\cos\frac{\pi}{2}t
+i x_{3,2}\sin\frac{\pi}{2}t )\,,\ \zeta_4(t)=\zeta_4\,;
\ 0\le t\le 1
\ee
(see Proposition 1.3 of \cite{FHIOPT}).
\setcounter{equation}{10}
\renewcommand{\theequation}{\thesection.\arabic{equation}}
The cross ratio $\i\,$ and the prefactor $D \equiv D(\uz ; p,p' )\,$
(\ref{D}) then change according to the law
\be
\lb{law}
z_{23} \to e^{-i\pi} z_{23}\quad \Rightarrow \quad
\i \to \frac{1}{\i}\,,\ \
D \to \bq^{\frac{n+1}{n}}\;\i^{\frac{{\frak p}+1}{h}}\; D\ \ \
\left( q^{\frac{1}{n}} =  e^{-i\frac{\pi}{nh}} \right)
\ee
where we have used (\ref{DD}) to derive the relations
\be
\lb{rlns}
\D (p)+\D (p') - 2\D' = \frac{\frak p}{h} - \frac{1}{nh}\,,\quad
2\D - \D_{(a)} = \frac{1}{h}+ \frac{1}{nh}\,.
\ee
The expansion (\ref{F=If}) into $SU(n)\,$ tensor invariants
changes under the combined action of analytic continuation along
the path ${\cal C}(2,3)\,$ and permutation of the indices $A,B\,$
as follows:
\ba
\lb{i1iF}
&&I_0^{AB}\,(1-\i )\,f^0_{\a\b}(\i) \,+\,
I_1^{AB}\,\i\, f^1_{\a\b}(\i)\,\ \stackrel{\curvearrowright}\to\\
&&\stackrel{\curvearrowright}\to\,
-\frac{1}{\i}\,\left( (I_0^{AB}\,+\, I_1^{AB})\,
(1-\i )\,f^0_{\a\b}(\frac{1}{\i})\, +\,
I_1^{AB}\, f^1_{\a\b}(\frac{1}{\i}) \right)\,.\nonumber
\ea
Inserting further the expansion (\ref{f=sS}) for $f^{\ell}_{\a\b}
(\i )\,,\ \ell = 0,1\,$ and, finally, using the relation
\ba
\lb{FFF}
&&F(\a ,\b ; \g ;\frac{1}{\i}) =
e^{-i\pi\a}\,\frac{\G(\g )\G(\b-\a)}{\G(\b )\G(\g-\a )}
\,\i^\a\,F(\a ,1+\a-\g ; 1+\a-\b ; \i ) +\nonumber\\
&&+ e^{-i\pi\b}\,\frac{\G(\g )\G(\a -\b )}{\G(\a )\G(\g-\b )}
\,\i^\b\,F(1+\b-\g , \b ; 1+\b-\a ; \i )\,,
\ea
we deduce (setting $k({\frak p}) := \frac{K_1}{K_0}\, )\,$
\be
\lb{B}
D\, s^{\ell}_\l (\i )\ \stackrel{\curvearrowright}\to\ D\, s^{\ell}_{{\l}'}
(\i )\, B^{{\l}'}_\l\,,\ \
B = \left(\matrix{B^0_0&B^0_1\cr B^1_0&B^1_1}\right)
= \bq^{\frac{1}{n}}\,\left(\matrix{
\frac{q^\fp}{[\fp ]}&k(\fp ) b_\fp \cr
{k(\fp )}^{-1} {b_{-\fp}}& -\frac{\bq^\fp}{[\fp ]}
}\right)
\ee
where $B\; ( \equiv B^s_2\, )\,$ is the {\em braid matrix}
corresponding to the exchange
$\stackrel{\curvearrowright}{2\,3}\,$ in the $s$-channel basis and
\ba
\lb{bp}
&&b_\fp\, =\, \frac{\G (1+\a-\b )\G (\a-\b)}{\G (2\a-\b )\G (1-\b )}\,
=\, \frac{\G (1+\frac{\fp}{h})\G(\frac{\fp}{h})}{\G(1+
\frac{\fp -1}{h})\G(\frac{\fp +1}{h})}\nonumber\\
&&\left(\
\Rightarrow \ \, b_{\fp} b_{-{\fp}} =
\frac{\sin\pi\frac{\fp +1}{h} \sin\pi\frac{\fp
-1}{h}}{\sin^2\pi\frac{\fp}{h}}=
\frac{[\fp +1][\fp -1]}{[\fp ]^2}\,
\right)\,.
\ea
Note that in line with the remark at the end of Section 2 the product
$q^{\frac{1}{n}}\, B\,$ of the braid matrix with a scalar phase
factor is independent of $n\,.$

As we already mentioned, the case $m=m'\,$ is much simpler.
Indeed, the space ${\cal F}(p, p' )\,$
(\ref{invt}), for $p\,, p+v^{(m)}\,$ and $p'=p+2v^{(m)}\,$
dominant, is one dimensional and so
is the space of quantum group invariants
(${\cal S}^0_{\a\b} = {\cal S}^1_{\a\b} =
q^{{\epsilon}_{\a\b}} {\cal S}^0_{\b\a} \equiv
{\cal S}_{\a\b}\,,$ cf. (\ref{f=sS}) and (3.3a)).
Due to the first equation (3.3b), the skewsymmetric $SU(n)\,$
invariant $I^{AB}_1\,$ is zero, see Appendix A,
and $I^{AB}_0 \equiv I^{AB}\,$ is symmetric, hence,
$\Omega_{23} I = 2 I\,,$ cf. (\ref{Omega}) and (\ref{I}).
The analogs of (\ref{w=DF}), (\ref{F=If}) and (\ref{f=sS}) read
then
\be
\lb{analogs}
w^{AB}_{\a\b} (\uz ; p,p' ) =
D(\uz ; p,p' )\, I^{AB}\, s (\i )\, {\cal S}_{\a\b}
\ee
and the KZ equation reduces to a first order equation for
$s (\i )\,:$
\be
\lb{KZeta1}
h \frac{d}{d\i} s (\i ) = - \frac{2}{1-\i}
s (\i )\,,\quad {\rm i.e.,}\quad
s(\i ) = K (1-\i )^{2\over h}\,.
\ee
Since in this case
\be
\lb{law1}
z_{23} \to e^{-i\pi} z_{23}\quad \Rightarrow \quad
1-\i \to e^{-i\pi}\; \frac{1-\i}{\i}\,,\ \
D \to \bq^{ \frac{n+1}{n} } \; \i^{ \frac{2}{h} }\; D \,,
\ee
we get simply
\be
\lb{i1is1}
D\, s (\i )\ \stackrel{\curvearrowright}\to\ q^{1-\frac{1}{n}}\,D\,
s (\i )\,.
\ee
All this fits perfectly the operator exchange relations,
\be
\lb{exch}
\stackrel{\curvearrowright}
{\v^B_i (z_3)\, \v^A_j (z_2)}\, =\, \v^A_{i'} (z_2)\, \v^B_{j'}
(z_3)
\, \R (\fp )^{i' j'}_{ij}
\ee
(related in \cite{FHIOPT} to the characteristic properties of the
intertwining quantum matrix algebra). In particular, the last
equation (3.3a) corresponds to the choice
\be
\lb{kp1}
k(\fp )\, =\,\frac{\G (\frac{1+\fp}{h}) \G (-\frac{\fp}{h}) }
{\G (\frac{1-\fp}{h}) \G (\frac{\fp}{h}) }\, \rho (\fp )\,,
\qquad \rho (\fp )\,\rho (- \fp )\, =1\, \left( \, =\, k (\fp )\,
k (- \fp )\,\right)
\ee
of the ratio of the normalization constants ${{K_1}\over{K_0}}\,.$
The resulting $4\times 4\,$ {\em dynamical
$R$-matrix} $\R (\fp )\,$ reads \cite{FHIOPT}
\ba
\lb{Rp4}
\R (\fp )\,&=&\,\left(\matrix{
\R (\fp )^{mm}_{mm}&\R (\fp )^{mm}_{mm'}&\R (\fp )^{mm}_{m' m}&\R (\fp )^{mm}_{m' m'}\cr
\R (\fp )^{mm'}_{mm}&\R (\fp )^{mm'}_{mm'}&\R (\fp )^{mm'}_{m' m}&\R (\fp )^{mm'}_{m' m'}\cr
\R (\fp )^{m' m}_{mm}&\R (\fp )^{m' m}_{mm'}&\R (\fp )^{m' m}_{m' m}&\R (\fp )^{m' m}_{m' m'}\cr
\R (\fp )^{m' m'}_{mm}&\R (\fp )^{m' m'}_{mm'}&\R (\fp )^{m' m'}_{m' m}&\R (\fp )^{m' m'}_{m' m'}
}\right) =\nonumber\\
&=&\bq^{1\over n}
\left(\matrix{
q&0&0&0\cr 0&\frac{q^\fp}{[\fp ]}&{{[\fp -1]}\over{[\fp ]}}
\rho (\fp )&0\cr 0&{{[\fp +1]}\over{[\fp ]}}\rho (-\fp )&
-\frac{\bq^\fp}{[\fp ]}&0\cr 0&0&0&q}\right) \,.
\ea
The braid matrices $B^s_1\,$ and $B^s_3\,$ corresponding to the
exchanges $\stackrel{\curvearrowright}{1\,2}\,$ and
$\stackrel{\curvearrowright}{3\,4}\,,$ respectively, are diagonal
with eigenvalues
\setcounter{equation}{0}
\renewcommand{\theequation}{\thesection.24\alph{equation}}
\be
\lb{B3}
\e_t\, q^{\D (p+v^{(t)}) - \D (p) - \D} \,=\,
\e_t q^{p_t -\frac{n-1}{2}}\,,\
\ee
($t=m,m'\,,\ \e_m=1\,,\ \e_{m'}=-1\,)\,$ -- for $B_3^s\,,$ and
\be
\lb{B1}
\e_t\, q^{\D (p+v^{(t)}) - \D (p' ) - \D} \,=\,
\e_t \bq^{p_r +\frac{n^2+n-4}{2n}}
\ee
$(p'=p+v^{(r)}+v^{(t)}=p+v^{(m)}+v^{(m')}\,)\,$ -- for $B_1^s\,.$
For $n=2$ and $p=(p_1 , p_2 ) = (1,-1) = p'\ \Leftrightarrow\
p_{12}=2={p'}_{12}\,$ - i.e., for the
four-point function of $4\,$ spin-$1/2$ operators, the braid
matrices $B^s_1\,$ and $B^s_3\,$ coincide:
\be
\lb{B13n=2}
B^s_1 = B^s_3 = \left(\matrix{q^{1\over 2}&0\cr 0& -\bq^{3\over
2}}
\right)\,.
\ee

\vspace{5mm}

\noindent
{\bf Remark 3.1~}
{The choice $\rho ({\frak p}) = \pm q^{\nu\frak p}
\sqrt{\frac{[{\frak p}+1]}{[{\frak p}-1]}}\,,\ \nu\,$ real, for the
remaining freedom in the normalization guarantees the unitarity
of $B\,$ (\ref{B}) (and of $\Rp\,$) for ${\frak p}+1<h\,,$
however, it violates the property of the elements of the matrix
$q^{\frac{1}{n}} B\,$ to belong to the cyclotomic field ${\Bbb Q} (q)\,$
which, as discussed in Section 5, could be useful.}

\vspace{5mm}

\section{Regular basis of solutions of the KZ equation. Triangular
braid matrices}

\medskip
\setcounter{equation}{0}
\renewcommand{\theequation}{\thesection.\arabic{equation}}

As noted in the Introduction, if we allow for unphysical values of
$p\,$ such that, say, $\fp = h\,,$ then the braid matrix $B_1\,$
(or $B_3\,$) is no longer diagonalizable, the corresponding solution
of the HG KZ equation having a logarithmic singularity.
We shall now introduce a regular basis of solutions which remains
meaningful also for such exceptional values of $\fp\,.$ To this
end we first introduce the $U_q(sl_n)\,$ counterpart of the basis
$I_0\,,\, I_1\,$ of $SU(n)\,$ invariants, setting
\be
\lb{qI}
{\cal I}^0_{\a\b} = {\cal S}^0_{\a\b} =
{\cal h} p' | a^m_\a a^{m'}_\b | p {\cal i}\quad (m<m')\,,\quad
{\cal I}^1_{\a\b} =
- {\cal I}^0_{{\a}' {\b}'} A^{{\a}' {\b}'}_{\a\b} =
{\cal I}^0_{\b\a} - q^{\epsilon_{\b\a}} {\cal I}^0_{\a\b}
\ee
where $A = (A^{{\a}' {\b}'}_{\a\b})\,$ is the quantum
antisymmetrizer of \cite{FHIOPT} satisfying $A^2 = [2] A\,$). We
shall set
\be
\lb{f=fJ}
f^{\ell}_{\a\b} (\i ) = \sum_{\l =0}^1 f^{\ell}_\l \,
{\cal I}^\l_{\a\b}\quad (= \sum_{\l =0}^1 s^{\ell}_\l\,
{\cal S}^\l_{\a\b}\ {\rm for}\ \fp <h)
\ee
(cf. (\ref{f=sS})), and will demonstrate that the solution of the
system (\ref{system}) is given by the contour integrals
\ba
\lb{contour}
&& f^{\ell}_0 (\i ) = N_0\,\int_{\i}^1 t^{\frac{\fp -1}{h} -\ell}
(1-t)^{\frac{1}{h}-1+\ell}(t-\i )^{\frac{1}{h}-1} dt =\nonumber\\
&&= N_0 B(\frac{1}{h} , \ell + \frac{1}{h} )
(1-\i )^{\frac{2}{h}-1+\ell} F(\ell - \frac{\fp -1}{h} ,
\ell + \frac{1}{h} ; \ell + \frac{2}{h} ; 1-\i )\,,\nonumber\\ \\
&&f^{\ell}_1 (\i ) = N_1\,\int^{\i}_0 t^{\frac{\fp -1}{h} -\ell}
(1-t)^{\frac{1}{h}-1+\ell}(\i -t)^{\frac{1}{h}-1} dt =\nonumber\\
&& = N_1 B(\frac{1}{h} , 1-\ell + \frac{\fp -1}{h} )
\i^{\frac{\fp}{h}-\ell}
F (1-\ell - \frac{1}{h} ,1-\ell + \frac{\fp -1}{h} ;
1-\ell + \frac{\fp}{h} ; \i )\nonumber
\ea
($\ell = 0,1\,$). The functions $f^{\ell}_\l (\i )\,$
are chosen in such
a way that if $B_2\,$ is the braid matrix associated with the
exchange $\stackrel{\curvearrowright}{2\,3}\,,$ then
\be
\lb{B2J}
( B_2 )^\l_\mu\ {\cal I}^\mu_{\a\b}\,=\,{\cal
I}^\mu_{{\a}'{\b}'}\, \R^{{\a}'{\b}'}_{\a\b}\,,\qquad
q^{\frac{1}{n}}\,\R^{{\a}'{\b}'}_{\a\b}\,=\,
q\,\d^{{\a}'}_\a \d^{{\b}'}_\b\,-\, A^{{\a}'{\b}'}_{\a\b}\,;
\ee
in other words, analytic continuation along the path
(3.11) combined with a permutation of the indices $A, B\,$ is
equivalent to the action of the constant (Jimbo) $R$-matrix.

Equating the expansion (\ref{f=fJ}) with (\ref{f=sS}) and using
(\ref{A.19}), (\ref{A.20}), we can find the relation with the
$s$-channel basis of KZ solutions:
\ba
\lb{relat}
&&s^{\ell}_0 (\i ) = f^{\ell}_0 (\i ) - \frac{[\fp -1]}{[\fp ]}
f^{\ell}_1\,,\qquad s^{\ell}_1 (\i ) =
\rho (\fp )\, \frac{[\fp -1]}{[\fp ]}
f^{\ell}_1 (\i )\nonumber\\
&&\quad\quad\left( \, \Rightarrow\ f^{\ell}_0 (\i)
= s^{\ell}_0 (\i ) + \rho (-\fp ) s^{\ell}_1 (\i )\ \right)\,.
\ea
While the integral representations (\ref{contour}) defining
$f^{\ell}_\l\,$ make perfect sense for all
${\fp} > \ell\,,$
the
amplitudes $s^{\ell}_\l\,$ are ill defined for ${\fp} = h\,.$ The
$\stackrel{\curvearrowright}{2\,3}\,$ braid matrix determined from
(\ref{B2J}) and (\ref{A.18}),
\be
\lb{B2-J}
B_2 = \bq^{\frac{1}{n}}\,\left(\matrix{q&1\cr 0& -\bq } \right)
\ee
agrees with (\ref{Rp4}). Indeed, it follows from
(\ref{Rp4}) and (\ref{A.19}) that
\ba
\lb{B2-J-S}
&&B_2 \left(\matrix{
1&0\cr -\frac{[\fp -1]}{[\fp ]}& \rho(\fp )\frac{[\fp -1]}{[\fp ]}
} \right) \left(\matrix{{\cal S}^0\cr{\cal S}^1} \right)
=\nonumber\\
&&= \left(\matrix{
1&0\cr -\frac{[\fp -1]}{[\fp ]}& \rho(\fp )\frac{[\fp -1]}{[\fp ]}
} \right)
\frac{\bq^{\frac{1}{n}}}{[\fp ]}
\left(\matrix{q^{\fp}&\rho(\fp )[\fp -1] \cr
\rho(-\fp ) [\fp +1]&-\bq^{\fp}}\right)
\left(\matrix{ {\cal S}^0\cr{\cal S}^1 } \right)\nonumber
\ea
which is consistent with (\ref{B2-J}) due to the $q$-number
identities
\be
\lb{qid}
q [\fp ] - [\fp -1] = q^{\fp}\,,\quad
[\fp +1] - q^{\fp} = \bq [\fp ]\,,\quad
[\fp -1] + {\bq}^{\fp} = \bq [\fp ]\,.
\ee

\vspace{5mm}

\noindent
{\bf Remark 4.1~} {
If we identify $\P_p\,$ in (\ref{w}) with another $SU(n)\,$ step
operator (setting $p_{12}=2\,,\ p_{i\, i+1}=1\,$ for $2\le i\le
n-1\,$), then we can speak of a monodromy representation of the
braid group ${\cal B}_3\,$ on 3 strands with generators
$B_2\,$ (\ref{B2-J}) and
\be
\lb{B3-J}
B_3 = \bq^{\frac{1}{n}}\,\left(\matrix{-\bq&0\cr 1& q } \right)
\ee
corresponding to the braiding
$\stackrel{\curvearrowright}{3\,4}\,:$
\ba
\lb{3-4}
&&z_{34}\ \rightarrow\ e^{-i\pi} z_{34}\,,\quad
\i\ \rightarrow\ \frac{e^{-i\pi}\i}{1-\i}\,,\quad
D \rightarrow\ \bq^{\frac{n+1}{n}} (1-\i )^{\frac{3}{h}}\, D\,,\\\nonumber
&&(1-\i ) I_0\ \rightarrow\ -\frac{1}{1-\i} I_0\,,\quad
\i I_1\ \rightarrow\ \frac{-\i}{1-\i} (I_0+I_1 )\,.
\ea
Note that in this case $p'\,$ determined by
${p'}_{12} = 2 = {p'}_{23}\,,\
{p'}_{i\, i+1} = 1\,$ for $3\le i\le n-1\,$
corresponds to the Young tableau
$
\put(0,2){\framebox(5,5)}
\put(0,-3){\framebox(5,5)}
\put(5,2){\framebox(5,5)}\quad
$
and hence is, in general, different from $p\,,$ except for $n=2\,$ when
$\P_{p'}\ (\, =\,\P_{{p'}^*}\, )\,$ is another step operator.
In the $n=2\,$ case
we are actually dealing with a special representation of the
braid group ${\cal B}_4\,,$ for which $B_1 = B_3\,.$
}
\vspace{5mm}

Eqs. (\ref{relat}), on the other hand, allow to relate the
normalization constants $K_\l\,$ of (\ref{s00})-(\ref{s11}) and $N_\l\,$
of (\ref{contour}). We find, comparing (\ref{s11}) with the second
equation (\ref{contour}) for $\ell =1\,$ and (\ref{relat})
\be
\lb{KN}
\frac{\fp}{\fp -1} K_1 = \rho (\fp )\frac{[\fp -1]}{[\fp ]}
N_1 B(\frac{1}{h} ,\frac{\fp -1}{h})\ \ \Rightarrow\ \ K_1 \,
=\, \rho(\fp ) \,N_1\, B(\frac{1}{h} , -\frac{\fp}{h})\,;
\ee
similarly, from (\ref{s00}), the first equation (\ref{contour}) and
(\ref{relat}) we deduce
\be
\lb{KN1}
K_0\, =\, N_0\, B(\frac{1}{h} ,\frac{\fp}{h})\quad\Rightarrow
\ \ k (\fp )\, =\, {{K_1}\over{K_0}}\, =\, \rho (\fp )\,
\frac{\G (-\frac{\fp}{h})\, \G (\frac{1+\fp}{h})}{\G (\frac{1-\fp}{h})\,\G
(\frac{\fp}{h})}\,\frac{N_1}{N_0}\,.
\ee
Comparing with (\ref{kp1}), we find
\be
\lb{NN}
N_0 \,=\, N_1\ \ \left(\; =
\frac{K_0}{B (\frac{1}{h} , \frac{\fp}{h} )}\,\right)\,.
\ee
Inserting (\ref{s00}), (\ref{s1}) and (\ref{contour}) into the last
equation (\ref{relat}) for $\ell = 0\,$
is equivalent to the following relation
for HG functions \cite{BE}:
\ba
\lb{BEHG}
&&\left(\frac{1}{N_0\, B(\frac{1}{h} , \frac{1}{h})}\, f^0_0 (\i )
=\right)\
(1-\i )^{\frac{2}{h} -1} F (\frac{1-\fp}{h} , \frac{1}{h} ;
\frac{2}{h} ; 1-\i ) =\nonumber\\
&&= \frac{\G (\frac{\fp}{h})\, \G (\frac{2}{h})}{\G (\frac{1}{h})\,\G (\frac{\fp + 1}{h})}
F (1- \frac{1}{h} , 1-\frac{\fp +1}{h} ;
1- \frac{\fp}{h} ; \i ) + \nonumber\\
&&+ \frac{\G (- \frac{\fp}{h})\, \G (\frac{2}{h})}{\G (\frac{1}{h})\,\G (\frac{1-\fp}{h})}
\,\i^{\frac{\fp}{h}}\,F (1- \frac{1}{h} , 1+\frac{\fp -1}{h} ;
1+ \frac{\fp}{h} ; \i )\ .
\ea
Note that the {\em poles} appearing for
$\fp = h\,$ in the right hand side of (\ref{BEHG}) cancel.

\vspace{5mm}

\section{Concluding remarks}

\medskip

The {\em regular basis} $f_\l\,$ (\ref{contour})
of solutions of the KZ equation (dual to the basis
${\cal I}^\l\,$ of $U_q(sl_n)\,$ invariants) is characterized by the
following properties.
\medskip

\noindent
(i)~~ The functions $f^\ell_\l\,,\ \ell = 0,1\,,$ are well defined for all
$\fp > \ell\,$ including the value $\fp = h\,$
(for which $s^\ell_\l\,$ blow up).
\medskip

\noindent
(ii)~ The braid matrices $B_2\,$ (\ref{B2-J}) and $B_3\,$
for $\fp = 2\,$ (see (\ref{B3-J})) are upper and lower
triangular, respectively. Unlike their $s$-channel basis counterparts
(as (\ref{B})), they have no singularities at $\fp = h\,.$ One
could say that the $s$-basis is ill defined since it pretends
to diagonalize the (non-diagonalizable for $\fp = h\,$) braid
matrix $B_3\,.$
\medskip

\noindent
(iii) The elements of the braid matrices $q^{\frac{1}{n}} B_a\,$ belong to the
cyclotomic field ${\Bbb Q} (q)\,.$ This remark has been exploited
in \cite{ST} to solve the Schwarz problem (of classifying the
cases when the monodromy representation of the braid group gives
rise to a finite matrix group). In fact, due to the observation
that the braid groups for ${\widehat{su}}(n)_k\,$ and
${\widehat{su}}(2)_{k+n-2}\,$ essentially coincide, the results of
\cite{ST} readily extend to the ${\widehat{su}}(n)\,$ case.

\vspace{5mm}

\noindent
L.H. and I.T. acknowledge partial support from the Bulgarian
National Council for Scientific Research under contract F-828.
The work of Ya.S. was supported in part by the EEC contract
HPRN-CT-2000-00122 and by the INTAS project 991590.

\vspace{5mm}

\section*{Appendix A. Bases and operators in the space of
$SU(n)\,$ and of $U_q(sl_n)\,$ invariant tensors}
\setcounter{equation}{0}
\def\theequation{A.\arabic{equation}}

\medskip

A basis in ${\cal F}(p , p')\,$ is conveniently expressed in terms
of an $SU(n)\,$ {\em intertwining matrix} $c= (c^A_i)\,$ whose entries
are subject to quadratic exchange relations
\be
\lb{A.1}
c^B_i c^A_j = c^A_{i'} c^B_{j'} \R^c (p)^{{i'}{j'}}_{ij}\,,\quad
p_i c^A_j = c^A_j (p_i + v_i^{(j)} )\,,\ \ v_i^{(j)} = \d^j_i -
\frac{1}{n}
\ee
and the determinant condition (in which the two $\e$-s
are totally antisymmetric for $r(\fp) = 1\,$ in (\ref{A.4}) below)
\be
\lb{A.2}
\e^{i_1 \dots i_n}\, c_{i_1}^{A_1}\dots c_{i_n}^{A_n} =
{\cal D}_1 (p)\, \e^{A_1 \dots A_n}\,,\quad
{\cal D}_1 (p) = \prod\limits_{1\le i<j\le n}\, p_{ij}\,.
\ee
The matrix $\R^c (p)\,$ satisfies the {\em dynamical Yang-Baxter
equation} \cite{GN, I2, HIOPT}
\ba
\lb{A.3}
&&\R^c_{12}(p) \R^c_{23}(p-v_1 ) \R^c_{12}(p) =
\R^c_{23}(p-v_1 ) \R^c_{12}(p) \R^c_{23}(p-v_1 )\nonumber\\
&&\qquad\qquad\left( \R^c_{23}(p-v_1 )^{i_1 i_2 i_3}_{j_1 j_2 j_3}
= \d^{i_1}_{j_1} \R^c (p-v^{(i_1 )})^{i_2 i_3}_{j_2 j_3} \right)
\ea
and the {\em ice condition} $\R^c (p)^{ij}_{kl}
= a^{ij}(p) \d^i_l \d^j_k + b^{ij}(p) \d^i_k \d^j_l\,.$
The solution (a special case -- for $q=1\,$ -- of the quantum
dynamical $R$-matrix given below)
depends on a non-zero function
$r (p_{ij})\,:$
\be
\lb{A.4}
\R^c (p)^{ii}_{ii}=1\ (\Leftrightarrow\, [c^A_i , c^B_i ] = 0 )\,,
\
\left(\matrix{
\R^c (p)^{mm'}_{mm'}& \R^c (p)^{mm'}_{m' m}\cr
\R^c (p)^{m' m}_{mm'}& \R^c (p)^{m' m}_{m' m}
}\right) = \frac{1}{\fp}
\left(\matrix{
1&\frac{\fp +1}{r(\fp)}\cr\frac{\fp -1}{r(-\fp)}&-1
}\right)
\ee
($m\ne m'\,,\ \fp = p_{m m'}\,$).
The involutivity of $\R^c (p)\,$ implied by (\ref{A.1}) fixes its
determinant to $-1\,:$
\be
\lb{A.5}
(\R^c (p) )^2 = \id\quad\Leftrightarrow\quad
r(\fp ) r(-\fp ) = 1\quad\Leftrightarrow\quad {\rm det} \R^c (p) =
-1\,.
\ee
We consider the Fock-type representation of the intertwining
matrix algebra with an $SU(n)$-invariant vacuum such that
\be
\lb{A.6}
c^A_i |0{\cal i} = 0\,,\ \ i>1\,,\quad
{\cal h} 0 | c^A_j = 0\,,\ \ j<n\quad (\,|0{\cal i} =
|p^{(0)}{\cal i}\,,\ p^{(0)}_{ij}=j-i\, )\,.
\ee
The meaning of the above relations is that $c^A_i\,,$ acting on a
ket vector $| p{\cal i}\,,$ adds a box to the $i$-th row of the
Young tableau associated with the $SU(n)\,$ IR of highest weight
$p\,$ (a more general statement, valid for $U_q = U_q(sl_n)\,$
and generic values of $q\,$ is proven in Section 3.1 of
\cite{FHIOPT}).

After these preliminaries we shall introduce (for $\fp > 1\,$) a basis
of eigenvectors of $\Omega_{12}\,$ (and hence, of $C_{12}\,$),
setting
\be
\lb{A.7}
S^{AB}_0 = T^{AB}_{m m'} := {\cal h} p' | c^A_m c^B_{m'} | p {\cal
i}\,,\quad
S^{AB}_1 = T^{AB}_{m' m} := {\cal h} p' | c^A_{m'} c^B_m | p {\cal
i}\,.
\ee
Indeed, for $p' =p+v^{(i)}+v^{(j)}\,$ we have
\be
\lb{A.8}
C_{12} T_{ij} = \frac{1}{2} \left( |p+v^{(j)}|^2 - | p' |^2 -
\frac{n^2-1}{n}\right)\, T_{ij} =
-\, (p_i + \d_{ij} + \frac{n^2+n-4}{2n} )\, T_{ij}\,.
\ee
Hence, in view of (\ref{Omega}),
\be
\lb{A.9}
\Omega_{12}\, S_0\, =\, 0\,,\quad
\Omega_{12}\, S_1\, =\, p_{m m'}\, S_1\quad
\Rightarrow\quad
\Omega_{12}^2 \,=\, p_{m m'}\,\Omega_{12}\,.
\ee
The spectrum of $\Omega_{23}\,,$ on the other hand, is found from
the last equation in (\ref{Omega}) which yields $\Omega_{23}^2 = 2
\Omega_{23}\,.$ We have
\be
\lb{A.10}
\Omega_{23} S_0 = (P_{23} + \id ) S_0\ \Rightarrow\
\Omega_{23} (P_{23} - \id ) S_0 = 0\,,\
\Omega_{23} (P_{23} + \id ) S_0 = 2 (P_{23} + \id ) S_0\,.
\ee
This suggests the introduction of the basis
(\ref{I}):
\be
\lb{A.11}
I_0 := S_0\,,\quad I_1  := (P_{23} - \id ) S_0\quad\left(\,
\Rightarrow\
P_{23} I_0 = I_0+I_1\,,\quad P_{23} I_1 = - I_1 \,\right)\,.
\ee
In order to express, conversely, $S_1\,$ in terms of $I_0\,,\,
I_1\,$ we use (\ref{A.1}), (\ref{A.4}) to derive
\be
\lb{A.12}
(P_{23} S_0 )^{AB} \,=\, T^{BA}_{m m'}\, =\, \frac{1}{\fp}\,
(\, T^{AB}_{mm'} + {r(\fp )} (\fp -1)\, T^{AB}_{m' m}\, )\,,
\ee
or
\be
\lb{A.13}
S^{AB}_1 ( \equiv T^{AB}_{m' m} ) = r(-\fp )
(\frac{\fp}{\fp -1} I_1^{AB} + I^{AB}_0 )\,,\quad
I^{AB}_1 =\frac{\fp -1}{\fp} (r(\fp ) S^{AB}_1 - S^{AB}_0 )\,.
\ee
The action of $\Omega_{ab}\,$ on $I_\ell\,$ is given by (\ref{I})
and by
\be
\lb{A.14}
\Omega_{23}\, I_0 \,=\, 2\, I_0\, +\, I_1\,,\quad
\Omega_{12}\, I_1 \,=\, \fp\, I_1\, + (\fp -1)\, I_0\,.
\ee
These relations imply also the last two equations (\ref{OOO}).

We proceed to displaying the interrelations between the $U_q\,$
invariants ${\cal S}^\l\,$ (\ref{f=sS}) and
${\cal I}^\l\,$ (\ref{qI}) derived from the properties of the
{\em quantum matrix algebra} generated by $a^i_\a\,$ and $q^{p_i}\,$
with basis exchange relations
\be
\lb{A.15}
\R (p)^{ij}_{i' j'}\, a^{i'}_\a\, a^{j'}_\b \,=\,
a^i_{{\a}'}\, a^j_{{\b}'}\, \R^{{\a}'{\b}'}_{\a\b}\,,\quad
q^{p_i}\, a^j_\a\,=\, a^j_\a\, q^{p_i + \d^j_i -\frac{1}{n}}
\ee
(plus a suitable determinant condition, see \cite{HIOPT, FHIOPT})
where $\R\,$ is the (Jimbo) $U_q\ n^2\times n^2\ R$-matrix
appearing in (\ref{B2J}).
The dynamical $R$-matrix $\R (p)\,$
again satisfies (\ref{A.3}) while the involutivity property
(\ref{A.5}) is replaced by a more general Hecke algebra condition
implying
\be
\lb{A.16}
q^{\frac{1}{n}}\, \R (p)\,=\, q\,\id \,-\, A(p)\,,\quad A (p)^2
\,=\,[2] \, A (p)\,.
\ee
The third equation in (3.3a) corresponds to the solution
\be
\lb{A.17}
A (p)^{ij}_{i' j'} = \frac{[p_{ij}-1]}{[p_{ij}]}
(\d^i_{i'} \d^j_{j'} - \rho (p_{ij}) \d^i_{j'} \d^j_{i'} )\,,\
i\ne j\,,\quad A^{ii}_{i' j'}=0=A^{ij}_{i' i'}
\ee
($\rho (p_{ij}) \rho (p_{ji}) = 1\,$).
It follows from the definition (\ref{qI}) that
\be
\lb{A.18}
q^{\frac{1}{n}}\, {\cal I}^0_{{\a}'{\b}'}
\,\R^{{\a}'{\b}'}_{\a\b}\, =
\, q\, {\cal I}^0_{\a\b}\, +\,{\cal I}^1_{\a\b} \,,\quad
q^{\frac{1}{n}}\, {\cal I}^1_{{\a}'{\b}'}
\,\R^{{\a}'{\b}'}_{\a\b}\,=\,
-\, \bq\ {\cal I}^1_{\a\b}\ .
\ee
Eqs. (\ref{f=sS}), (\ref{qI}), (\ref{B2J}) and
(\ref{A.15})-(\ref{A.17}), on the other
hand, imply
\be
\lb{A.19}
{\cal I}^1_{\a\b} \, =\, \frac{[\fp -1]}{[\fp ]}\,
\left(\,
\rho (\fp )\, {\cal S}^1_{\a\b}\,-\,{\cal S}^0_{\a\b}
\right)\,,\quad \fp = p_{m m'}\quad (m\ne m' )\,.
\ee
Conversely, ${\cal S}^\l_{\a\b}\,$ are expressed in terms of
${\cal I}^\l_{\a\b}\,$ by
\be
\lb{A.20}
{\cal S}^0_{\a\b}\, =\, {\cal I}^0_{\a\b}\,,\quad
\rho (\fp )\, {\cal S}^1_{\a\b}\, =\, {\cal I}^0_{\a\b}\, +\,
\frac{[\fp ]}{[\fp -1]}\,{\cal I}^1_{\a\b}\,.
\ee
As noted above, relations (\ref{A.11}), (\ref{A.13})
appear as the $q \to 1\,$ limit of (\ref{A.19}), (\ref{A.20})
(for ${\cal I}^0 \, ( = {\cal S}^0\, ) \,\to I_0\, ( = S_0\, )\,,$ resp.
${\cal I}^1 \to I_1\,,\ {\cal S}^1 \to S_1\,,$ 
and
$\rho (\fp ) \to r (\fp )$).

\vspace{5mm}


\end{document}